\documentclass{article}

\usepackage{microtype}
\usepackage{graphicx}
\usepackage{subfigure}
\usepackage{booktabs} 

\usepackage{hyperref}

\usepackage[accepted]{icml2025}

\usepackage{amsmath}
\usepackage{amssymb}
\usepackage{mathtools}
\usepackage{amsthm}

\usepackage[capitalize,noabbrev]{cleveref}

\theoremstyle{plain}

\theoremstyle{definition}

\theoremstyle{remark}

\usepackage[textsize=tiny]{todonotes}

\begin{document}

\twocolumn[
\icmltitle{Bitstream Collisions in Neural Image Compression via Adversarial Perturbations}



\icmlsetsymbol{equal}{*}

\begin{icmlauthorlist}
\icmlauthor{Jordan Madden}{equal,yyy}
\icmlauthor{Lhamo Dorje}{equal,yyy}
\icmlauthor{Xiaohua Li}{equal,yyy}
\end{icmlauthorlist}

\icmlaffiliation{yyy}{Department of ECE, Binghamton University, Binghamton, NY, USA 13902. 
 \{jmadden2,ldorje1,xli\}@binghamton.edu}

\icmlcorrespondingauthor{Xiaohua Li}{xli@binghamton.edu}

\icmlkeywords{Neural Image Compression, Adversarial Machine Learning, Collision, Defense, Robustness}

\vskip 0.3in
]



\printAffiliationsAndNotice{} 






\begin{abstract}
Neural image compression (NIC) has emerged as a promising alternative to classical compression techniques, offering improved compression ratios. Despite its progress towards standardization and practical deployment, there has been minimal exploration into it's robustness and security. This study reveals an unexpected vulnerability in NIC - bitstream collisions - where semantically different images produce identical compressed bitstreams.  Utilizing a novel whitebox adversarial attack algorithm, this paper demonstrates that adding carefully crafted perturbations to semantically different images can cause their compressed bitstreams to collide exactly. 
The collision vulnerability poses a threat to the practical usability of NIC, particularly in security-critical applications. 
The cause of the collision is analyzed, and a simple yet effective mitigation method is presented. 
\end{abstract}

\section{Introduction}

Image compression has become crucial in today's big data regime as it enables efficient storage and transmission of image data while maintaining visual quality. Many novel media technologies such as Virtual Reality (VR), Augmented Reality (AR), Extended Reality (ER), etc, have raw data rates that are orders of magnitude above the wireless communication capacity bound, making efficient image compression especially important \cite{bastug2017toward}.

In recent years, Neural Image Compression (NIC) has become popular due to its ability to outperform traditional compression techniques like JPEG (Joint Photographic Experts Group) or PNG (Portable Network Graphics) \cite{ yang2023introduction, jamil2023learning}. By leveraging the power of deep learning, NIC can optimize compression end-to-end, achieving higher compression ratios while maintaining high perceptual quality.

Despite the fact that NIC algorithms are expected to be standardized and implemented in practical applications - even security-critical ones such as cryptographic protocols - their robustness and security has been largely unexplored \cite{liu2023manipulation, chen2023toward}. Traditional compression methods like JPEG have undergone extensive testing before real-world adoption, resulting in well-understood vulnerabilities.



\begin{figure}[t]
\centerline{\includegraphics[width=8.5cm]{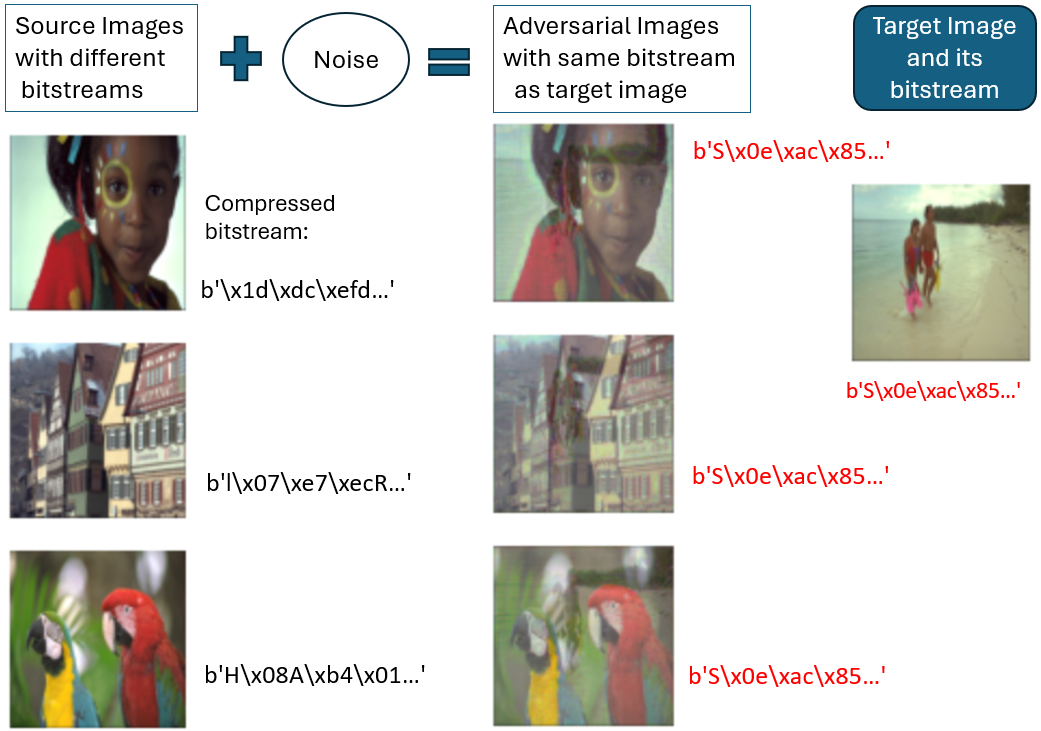}}
\caption{Bitstream collisions: Adversarial images that are perceptually different from a target image but have the same compressed bitstream as the latter.}

\label{fig1sampleimages}
\end{figure}

In this paper, we show that the NIC suffers from an unexpected vulnerability, i.e., bitstream collisions, where semantically different images can be compressed to the same bitstream.  Some sample collision images that we derived are illustrated in Fig. \ref{fig1sampleimages}. For any target image $x_{tgt}$ with compressed bitstream $b_{tgt}$, we can find adversarial images $x_{adv}$ that are semantically different from $x_{tgt}$ but compressed to the same bitstream $b_{tgt}$. The image $x_{adv}$ can be perceptually similar to any arbitrary source image $x_{src}$. 
 
For traditional codecs like JPEG, it is well understood that if two images produce similar compressed bitstreams, the images are likely perceptually similar, with minor differences in noise or high-frequency variations. This paper shows that no such guarantee is available for  NIC. Since semantically different images may be compressed to identical bitstreams, this exposes a severe vulnerability in many applications. It is ambiguous at the decompressor side which image, $x_{adv}$ or $x_{tgt}$, is the true uncompressed image. For example,  in video surveillance or face recognition, the image is compressed and transmitted to some central processor for recognition. The collisions may lead to incorrect recognitions. 

\begin{figure}[t]
\centerline{\includegraphics[width=8cm]{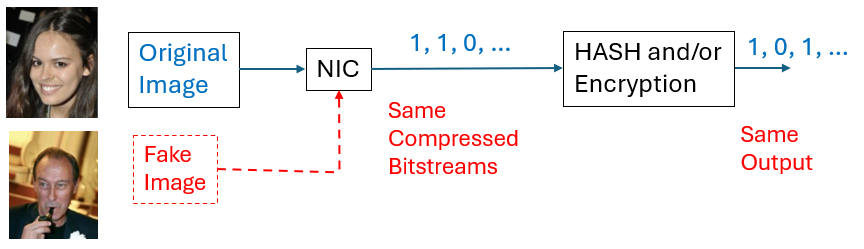}}
\caption{NIC with a collision vulnerability compromises cryptographic protocols.}
\label{fig2crypto}
\end{figure}

The most detrimental consequence is to cryptographic protocols that integrate NIC, as illustrated in Fig. \ref{fig2crypto}. Cryptographic protocols used in digital signatures, digital IDs, blockchains, etc, may compress an image and then hash/encrypt the compressed bitstream \cite{menezes2018handbook, gelb2018identification}. If NIC is applied, both the integrity and non-repudiation properties of the cryptographic protocols will be compromised due to NIC's collision vulnerability. An attacker may replace the original image with fake images but still obtain the same hash/encryption output. The attacker may also argue that the image used in hash or encryption is not the original one.



This paper investigates the NIC collision vulnerability with the following contributions:

\begin{itemize}
    \item We develop a Masked Gradient Descent (MGD) attack algorithm to generate collision images. To the best of our knowledge, this is the first time  the collision vulnerability of NIC is discovered and demonstrated.
    
    \item We analyze the collision problem theoretically and propose a novel Limited Precision Defense (LPD) method that is simple to implement and can effectively mitigate the collision vulnerability.
    
    \item We evaluate the efficacy of the proposed attack and defense across multiple datasets, compression quality factors, and NIC model architectures.
    
\end{itemize}

The paper is organized as follows: Section \ref{literature-review} provides a literature review.
Section \ref{attack} formulates the new attack algorithm and defense method. Section \ref{experiment} presents the experiment results, and Section \ref{conclusion} concludes the paper. 
\footnote{Our source code can be found at \url{https://github.com/neddamj/nicsec}.}

\section{Literature Review}
\label{literature-review}
 
\subsection{Neural Image Compression}\label{AA}

Using neural networks for image compression has attracted long-term research interests \cite{dony1995neural, cramer1998neural}. Surveys of NIC as well as the current state-of-the-art can be found in \cite{ma2019image, yang2023introduction, jamil2023learning}.

  In the initial stage of deep learning-based NIC development, works  dealt with the issues of non-differentiable quantization and rate estimation to enable end-to-end training of NIC models \cite{agustsson2017soft, balle2016end2end}. Subsequent research shifted towards optimizing network architectures for efficient latent representation extraction and high-quality image reconstruction \cite{yang2020improving, guo2021soft, yang2021slimmable, zhu2022unified, gao2022flexible, strumpler2022implicit, muckley2023improving}. Recurrent neural networks, as employed in works such as \cite{toderici2017full, toderici2015variable} demonstrated success in compressing residual information recursively.  Generative models were used to learn the image distributions, achieving improved subjective quality at very low bit rates, as seen in \cite{rippel2017real, santurkar2018generative, mentzer2020high}. Additionally, techniques such as \cite{balle2018variational, minnen2018joint, cheng2020learned} focused on adaptive context models for entropy estimation to achieve the optimal tradeoff between reconstruction error and entropy. 

\subsection{Adversarial Machine Learning}

Adversarial attacks can be classified as either untargeted or targeted. In an untargeted attack, the attacker aims to slightly change the input to the model, so that the model produces a wildly different output. In a targeted attack, the attacker changes the input to the model slightly to cause the model to generate a predefined output. Adversarial attacks can also be categorized based on the level of access the attacker has to the target model. In whitebox attacks \cite{szegedy2013intriguing, madry2017towards, carlini2017towards}, the attacker has complete knowledge of the model, including its architecture and parameters. On the other hand, blackbox attacks \cite{ilyas2018black, liu2019signsgd, brendel2017decision, chen2020hopskipjumpattack} occur when the attacker has no access to the model’s inner workings, relying only on inputs and outputs. Blackbox attacks can be further divided into two types: soft-label attacks \cite{ilyas2018black, liu2019signsgd}, where the model outputs continuous values like probabilities or logits, and hard-label attacks \cite{brendel2017decision, chen2020hopskipjumpattack}, where the model provides only discrete decision labels. 
In this work, we will develop new targeted whitebox attacks to create collision images. 

\begin{figure*}[t]
\centerline{\includegraphics[width=14cm]{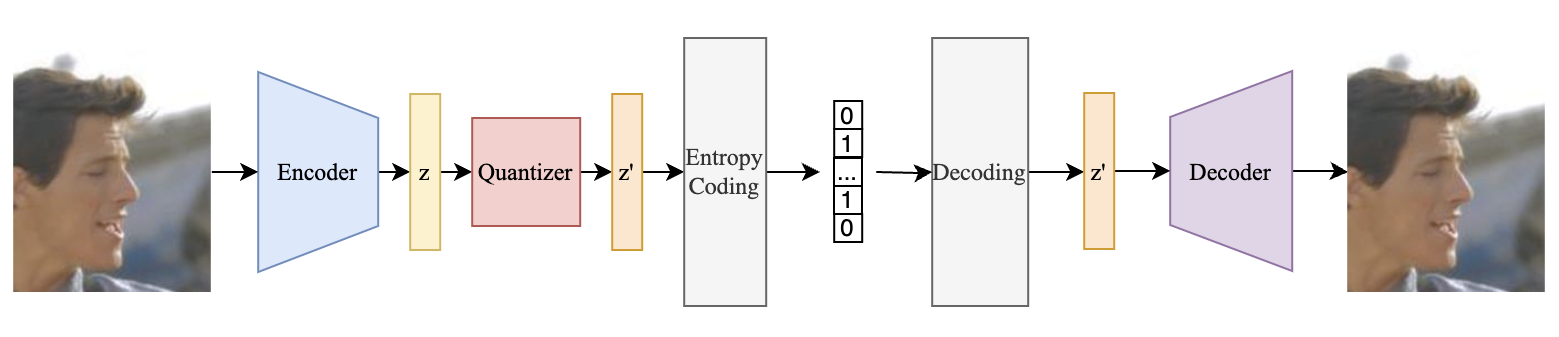}}
\caption{Standard Neural Image Compression and Decompression Pipeline}
\label{pipeline}
\end{figure*}

\subsection{Adversarial Robustness of NIC}


In contrast to the abundant literature on either NIC or adversarial machine learning, there is not much work done on NIC robustness. \cite{liu2023manipulation} demonstrated that the introduction of adversarial noise can lead to a substantial increase in the compressed bit rate. Additionally, \cite{chen2023toward} discovered that even the injection of minuscule amounts of adversarial noise can result in severe distortions in the reconstructed images. With the introduction of JPEG AI \cite{ascenso2023jpeg}, there is now a standard for end-to-end JPEG compression methods. In response, \cite{kovalev2024exploring} developed a novel methodology specifically designed to assess the robustness of JPEG models against adversarial attacks. 

As far as we know, this work is the first to study the compressed bitstream collision problem. 


\section{Robustness of Neural Compression}
\label{attack}

\subsection{NIC Model} \label{nicmodel}

We consider lossy compression as described in \cite{yang2023introduction}. The NIC models are inspired by the model of transform coding, in which an analysis transform $f$ and a synthesis transform $g$ are jointly optimized to compress/decompress an image with certain rate-distortion performance. As shown in Fig. \ref{pipeline}, for an image $x$, $f$ computes a continuous representation $z = f(x)$, which is then quantized to ${z}' = \lfloor z \rceil$. Entropy coding is applied to create the compressed bitstream $b=e(z')$. The decoding process reverts the entropy coding to get $z'$ from $b$, and the synthesis transform is subsequently applied to reconstruct the original image as $x' = g(z')$.

In traditional compression, the analysis and synthesis are conducted by orthogonal linear transformations such as discrete Fourier transform.
NIC uses deep neural networks (DNNs) instead, where the analysis and synthesis transforms are typically replaced by  encoder and decoder models based on Convolutional Neural Networks (CNNs) \cite{theis2017lossy, balle2016end}.
The various components of the pipeline in Fig. \ref{pipeline} are jointly optimized over a rate-distortion objective, such as $\mathbb{E}[-\log_2 p(\lfloor f(x) \rceil)] + \lambda \mathbb{E}[{\cal D}(x, g(\lfloor f(x)\rceil))]$,
where $\mathbb{E}$ is expectation, $\lambda$ is the trade-off factor, $p(\lfloor f(x)\rceil)$ measures the bit rate, and ${\cal D}$ is a distance metric that measures the image distortion. 


\subsection{Threat Model}
Consider a NIC model and an image $x_{tgt}$ with its compressed bitstream $b_{tgt}$. The attacker's objective is to create an image $x_{adv}$ that is semantically different from $x_{tgt}$ but is also compressed to the same bitstream $b_{tgt}$.  We assume that the attacker has whitebox access to the NIC model, allowing full knowledge of the model architecture, parameters, and gradients. Additionally, $x_{adv}$ should look like a normal image with added perturbations small enough. To guarantee this, one approach is for the attacker to select an image $x_{src}$ that is completely different from $x_{tgt}$ and gradually perturb the former so that its compressed bitstream converges to the latter's bitstream. This is similar to the methodology of some targeted blackbox attacks \cite{ilyas2018black}.


\subsection{Attacker's Masked Gradient Descent (MGD) Algorithm for Collision Generation}

Starting from an image $x_{src}$, the attacker's goal is to create $x_{adv}$ so that it has the same bitstream as $x_{tgt}$, specifically, $b_{adv} = b_{tgt}$, or  $e(\lfloor f(x_{src}) \rceil) = e(\lfloor f(x_{tgt}) \rceil)$. This could be achieved via the optimization
\begin{equation}
  x_{adv} = {\rm arg} \; \min_{x} \; \left\| e(\lfloor f(x) \rceil) - e(\lfloor f(x_{tgt}) \rceil) \right\|,  \label{eq1}
\end{equation}
where $\| \cdot \|$ can be some norm or Hamming distance, and $x$ is initialized as $x_{src}$ to ensure $x_{adv}$ is semantically different from $x_{tgt}$. Unfortunately, the quantization and entropy encoding procedures make the gradient optimization of (\ref{eq1}) challenging, if not impossible. 

Observing that the quantization and entropy encoding used in NICs are conventional signal processing tasks that discard very small signal variations only, we propose to skip them and directly minimize the distance between the CNN encoder's output logits $f(x_{adv})$ and $f(x_{tgt})$, which gives
\begin{align}
 x_{adv} ={\rm arg} \; \min_{x} \; \left\| f(x) - f(x_{tgt}) \right\|.   \label{eq10}
\end{align}
If the optimized loss ${\cal L}(x_{adv}) = \left\| f(x_{adv}) - f(x_{tgt}) \right\|$ is small enough, the bitstreams of $x_{adv}$ and $x_{tgt}$ will become identical.

The new challenge is that optimizing (\ref{eq10}) directly using gradient descent algorithm, or variations like Projected Gradient Descent (PGD) \cite{madry2017towards} or the algorithm of \cite{carlini2017towards}, largely fails. They either do not converge to a loss small enough for bitstream collision, or converge to $x_{adv}\approx x_{tgt}$, i.e., the adversarial image becomes just the target image. Through experiments, we find that when a patch of the adversarial image $x_{adv}$ was edited, a corresponding, similarly sized patch of the decompressed adversarial image $g(\lfloor f(x_{adv}) \rceil)$ was also changed. This indicated that each pixel that was changed in $x_{adv}$ affected the corresponding pixel in $g(\lfloor f(x_{adv}) \rceil)$, as well as a few pixels in its immediate vicinity. This meant that we could not freely run optimization on the entire  image $x_{adv}$ as the resulting perturbation would cause $x_{adv}$ to look exactly like $x_{tgt}$. 

Surprisingly, we find that the challenge can be resolved by applying a simple masking technique to the gradients.  The role of the mask is just to set many gradient elements to zero before using the gradient to update the image. This allows the majority of the pixels in the image $x_{adv}$ to remain unchanged. Only some pixels in the image are allowed to be perturbed.

Specifically, we use what we call a \textit{dot mask} where a fixed portion of pixels, separated by a certain rows and columns, are allowed to be perturbed by the gradient. The dot mask function is formulated as 
\begin{equation}
   {\cal M}(g(h, w)) = \left\{ \begin{array}{ll}
       g(h, w),  &  {\rm for} \; h=i\Delta_h + h_0, \\
       &    \;\;\;\;\; w=i\Delta_w + w_0, \\
       &    \;\;\;\;\;\; i = 0, 1, \cdots ; \\
       0,  & {\rm else} 
       \end{array} \right.    \label{eq15}
\end{equation}
where $g(h, w)$ is the gradient of the image $x_{adv}$ at height $h$ and width $w$, $\Delta_h$ and $\Delta_w$ are the stepsizes or strides across the height and width, whereas $h_0$ and $w_0$ are initial shiftings. The optimization of (\ref{eq10}) can thus be conducted by the masked gradient (\ref{eq15}) as follows:
\begin{equation}
    x \leftarrow x - \mu {\cal M}\left( \frac{\partial \| f(x) - f(x_{tgt}) \|} { \partial x}  \right)   \label{eq20}
\end{equation}
where ${\cal L}'(x) = \partial \| f(x) - f(x_{tgt}) \| / \partial x$ is the gradient of the loss with respect to $x$. 

The new attack algorithm is outlined in 
Algorithm \ref{pseudo_algo}. Because our main objective is to generate collision images, not enhance image quality, we do not apply any constraints over image distortions. Instead, the image distortion is indirectly suppressed via the early-stopping technique. The iterative optimization stops immediately whenever the bitstream collision occurs. 

\begin{algorithm}[t]
\caption{MGD: Masked Gradient Descent Algorithm}
\textbf{Initialization}: Given total iteration $I$, $x_{tgt}$ and its bitstream $b_{tgt}$, initialize $x = x_{src}$. 

\textbf{For $i = 0, 1, \cdots, I-1$, do}

\begin{itemize}
\item Update (\ref{eq20}) with respect to (\ref{eq10}) and (\ref{eq15}),
\item Compare the new $x$'s compressed bitstream $b$ with  $b_{tgt}$. Stop and output $x_{adv}=x$ if $b=b_{tgt}$.
\end{itemize}

\label{pseudo_algo}
\end{algorithm}

One may suspect that reducing $\Delta_h$ and $\Delta_w$ to 1 would improve the attack's effectiveness by allowing the adversarial perturbation to induce more change in the image. However, it would cause $x_{adv}$ to converge to $x_{tgt}$ instead of a perceptually different image. Therefore, it is critical to choose appropriate mask parameters as it helps to sufficiently maintain the semantic properties of the image $x_{src}$ while allowing the attack to work sufficiently well.
In practice, we find the attack to work best when we separate the dots horizontally by $\Delta_w=1$ pixel and vertically by $\Delta_h=3$ pixels. The shiftings can be simply set as $h_0=w_0=0$.


Although we use a dot mask, any mask that distributes the perturbations throughout the image such that the image retains its semantic properties and the output converges to that of the target image would work. In addition, any loss functions that reduce the difference between $f(x_{adv})$ and $f(x_{tgt})$ would also work under the mask operation. We tried both mean square error and Cosine Similarity (CS), and found either of them, or their combination such as the following, worked well
\begin{align}
{\cal L}(x) = \left\| f(x) - f(x_{tgt}) \right\| + \frac{1}{2}\left[1 - CS(x, \; x_{tgt}) \right]  
\end{align}
where $CS(a, b) = \frac{a \cdot b} {\| a \| \| b \|}$.

\subsection{NIC's Defense: A Limited-Precision Defense (LPD) Mechanism}

Various common defense mechanisms applied in robust machine learning can be explored to mitigate NIC's collision vulnerability, such as the adversarial training used in \cite{chen2023toward}. 
Nevertheless, we are more interested in the following NIC-specific mechanism, which we call {\it Limited-Precision Defense (LPD)}. It exploits a special characteristic of NIC, i.e., sensitivity to small perturbations. It is much simpler to implement and is also very effective.

In our initial experiments, we observed a surprising phenomenon: for the attack to work, we needed to disable the TF32 tensor cores on our GPUs; otherwise we could not get any collisions. Through careful study, we found that the reduced precision of TF32 interfered with the convergence of  Algorithm 1. Even though the loss ${\cal L}(x)$ can still be as small as in the case without using TF32, the compressed bitstream of the adversarial image could hardly be equal to that of the target image. The reason is that the compressed bitstream is extremely sensitive to even minor variations in the latent representations. 

Drawing from this observation, we propose the unique {\it LPD} method, i.e., converting some latent tensors and/or some model parameters to half-precision. If a NIC model is developed with float32, then we change some data to float16. It turns out that this simple change can effectively mitigate the attack algorithms. 

This method can be directly applied in real NIC applications. Especially for the crypto application shown in Fig. \ref{fig2crypto}, the limited-precision implementation of NIC does not affect the normal operation of the crypto-chain. Each unique image will still lead to a fixed and unique compressed bitstream. The adversary, or anyone else, can recreate this compressed bitstream using the same limited-precision implementation. However, if high-precision implementation is used, the generated bitstream will be different from the true one. Without a high-precision implementation, the attack algorithms can hardly converge to valid collision images. 


\subsection{Theoretical Bounds of the Distance between Collision Images}

To be considered as a vulnerability, collision images must be semantically different, or the distance between the images must be large. In this section, we derive the bounds of such distances between collision images. 


Recall the encoding function $f$ (see Section \ref{nicmodel} and Fig. \ref{pipeline}) that transforms an image $x$ to their latent embeddings (or logits)  $z = f(x)$. We skip the details of the quantization and entropy encoding, but model the data compression procedure as follows. Let 
\begin{align}
    z = z_{b} + z_{v}     \label{3.10}
\end{align}
where $z_{b}$ is a vector consisting of all the values of $z$ whose magnitudes are larger than a threshold $\gamma$, while $z_{v}$ is a vector consisting of all the values with magnitude less than the threshold $\gamma$. The data compression is to discard $z_{v}$ while keeping just $z_{b}$. 
With this compression model, the compression ratio $R$, defined as the ratio between the total number of elements in $z$ and the number of elements in $z_{b}$, becomes
\begin{equation}
   R = \frac{1}{\mathbb{P}[|z| > \gamma]} =\frac{1}{2 \int_\gamma ^\infty p(z) dz}  \label{3.20}
\end{equation}
where $p(z)$ is the distribution of $z$.

{\bf Theorem 1}. {\it Assume the elements of images $x_{tgt}$ and $x_{adv}$ are independent with identical normal distribution ${\cal N}(0, 1)$. In conventional compression with orthogonal transform $f$, if $x_{tgt}$ and $x_{adv}$ are compressed to the same bitstream, then their per-pixel distance, defined as
${\cal D}_c (x_{tgt}, x_{adv}) \stackrel{\triangle}{=} \
\left\{\mathbb{E}[\| x_{tgt} - x_{adv} \|^2/M]\right\}^{1/2}$ with $L_2$ norm and total $M$ pixels, is bounded by $\sqrt{2}$, i.e.
\begin{equation}
   {\cal D}_c (x_{tgt}, x_{adv}) \leq  \sqrt{2}. \label{eq3.30}
\end{equation}
}

{\it Proof}:
If $f$ is an orthogonal transform such as discrete Fourier transform (DFT) or singular value decomposition (SVD), then the elements of $z_{tgt}=f(x_{tgt})$ and $z_{adv}=f(x_{adv})$ are also distributed as ${\cal N}(0, 1)$.  From (\ref{3.10}), we have $z_{adv}=z_{adv, b}+z_{adv,v}$ and $z_{tgt} = z_{tgt, b}+z_{tgt, v}$. We also have $z_{adv,b}=z_{tgt,b}$ because they are compressed to the same bitstream.
The distance square between the two images $x_{adv}$ and $x_{tgt}$ can be deduced as
\begin{align}
{\cal D}_c^2(x_{tgt}, x_{adv}) 
  &= \mathbb{E}\left[ \| z_{tgt} - z_{adv} \|^2/M \right]     \nonumber \\
  &= \mathbb{E}\left[ \| z_{tgt, v} - z_{adv, v} \|^2/M \right] 
  \label{3.40}
\end{align}
Because all the elements of $z_{tgt, v}$ and $z_{adv, v}$ (each of them has $M(1-1/R)$ elements) are less than $\gamma$, we have
\begin{equation}
    {\cal D}_c^2(x_{tgt}, x_{adv}) =  \frac{R-1}{R}\int_{-\gamma}^\gamma \int_{-\gamma}^\gamma (x-y)^2 p(x)p(y)dxdy   \label{eq3.50}
\end{equation}
where both $p(x)$ and $p(y)$ are ${\cal N}(0, 1)$.
Next, to prove (\ref{eq3.30}), we use the property that ${\cal D}_c$ increases monotonically with $\gamma$. Let $\gamma = \infty$, we can calculate (\ref{eq3.50}) to be $2$.   
\hfill{$\Box$}

\begin{figure}[t]
\centerline{\includegraphics[height=6cm]{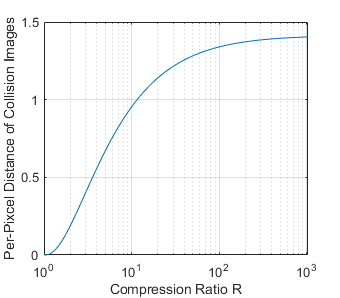}}
\caption{Theoretical limit of collision image distance as a function of compression ratio for conventional compressors with orthogonal $f$.}
\label{nictheory1}
\end{figure}

We have verified Theorem 1 via simulations. We derived closed-form solutions to (\ref{3.20}) and (\ref{eq3.50}), and plotted the distance curve in Fig. \ref{nictheory1}.
It can be seen that the distance increases with the compression ratio up to the bound $\sqrt{2}$.

For NIC, although accurate distance is unavailable, we can still derive useful information about the bounds. 

{\bf Theorem 2.} {\it Different from conventional deep model robustness that requires a small Lipschitz constant, large Lipschitz constant is helpful to collision mitigation.}

{\it Proof}. Assume $f$ is a deep model that is Lipschitz continuous \cite{fazlyab2019efficient} with Lipschitz constant $L$, i.e.
$\| f(x_{tgt}) - f(x_{adv})\| \leq L \| x_{tgt} - x_{adv} \|$.
Following (\ref{3.40}), the distance square can be calculated as follows
\begin{align}
   {\cal D}_n^2(x_{tgt}, x_{adv}) &= \mathbb{E}\left[\| x_{tgt} - x_{adv} \|^2/M\right] \nonumber \\
  & \geq \frac{1}{L^2} \mathbb{E}\left[ \| f(x_{tgt}) - f(x_{adv)} \|^2/M \right]  \nonumber \\
  &= \frac{1}{L^2} \mathbb{E}\left[ \| z_{tgt, v} - z_{adv, v} \|^2/M \right].  \label{eq3.80}
\end{align}
This means that the lower bound of the distance is decreased by $L$. Increasing $L$ can make the NIC more robust to collisions, which is different from conventional deep model robustness requirements.  \hfill{$\Box$}

The upper bound of the distance is more interesting than the lower bound. To derive the upper bound, instead of the Lipschtz constant, we need to consider the following new constant $C$ which we call {\it contraction constant}
\begin{equation}
   C = \max_{x, y} \frac{ \|x-y\| } {\| f(x) - f(y) \|}.
   \label{eq3.90}
\end{equation}

{\bf Theorem 3}. {\it Assume $f$'s output is normalized to ${\cal N}(0, 1)$ in NICs, then ${\cal D}_n(x_{tgt}, x_{adv}) \leq \sqrt{2}C$.
}

{\it Proof}. The definition (\ref{eq3.90}) means 
\begin{equation}
    \| x_{tgt} - x_{adv} \| \leq C \| f(x_{tgt}) - f(x_{adv}) \|.
\end{equation}
Then, with a similar procedure as (\ref{eq3.80}), we can get
\begin{align}
   {\cal D}_n^2 (x_{tgt}, x_{adv}) \leq C^2 \mathbb{E}\left[ \| z_{tgt, v} - z_{adv, v} \|^2/M \right].
\end{align}
Following (\ref{eq3.50}), we can easily prove this theorem.  \hfill{$\Box$}

Theorem 3 shows that the upper bound of the distance is $C$ times higher in NICs than in conventional compressors (\ref{eq3.50}).
For more robust NIC, we need to look for $f$ with a small $C$. This can be a new mitigation method to explore in the future.


\section{Experimental Results and Discussion} \label{experiment}

\subsection{Experiment Settings}

We evaluated our proposed {\bf MGD} attack algorithm and the {\bf LPD} defense method using three distinct datasets: CelebA, ImageNet, and the Kodak dataset, with all images resized to $256 \times 256$ pixels. Additionally, we compared the proposed MGD attack algorithm against two standard attack algorithms: the Projected Gradient Descent ({\bf PGD}) attack algorithm \cite{madry2017towards} and the Carlini-Wagner ({\bf CW}) attack algorithm \cite{carlini2017towards}.

Most neural compression algorithms are significantly more computationally intensive than traditional codecs and this may hinder their adoption in practice. As a result, our work emphasizes compressors that deliver strong performance while utilizing fewer parameters. In our experiments, we adopted three NIC models that satisfy this criterion: two Factorized Prior models ({\bf FP-GDN} and {\bf FP-ReLU}) and the Scale Hyperprior model ({\bf SH}) presented in \cite{balle2018variational}.
For Factorized Prior models, we used two implementations: the model using the Generalized Divisive Normalization (GDN) layers \cite{balle2016density} and the one using ReLU layers instead of GDN layers. For comparison, we also evaluated the robustness of JPEG under the above attacks. 


For the Factorized Prior models, we ran attack optimization for at most $I=5,000$ iterations, whereas for the Scale Hyperprior model, we ran optimization for at most $I=20,000$ iterations. In both cases, we used the Adam optimizer \cite{kingma2014adam} with an initial learning rate of 0.03 to perform optimization. We varied the learning rate throughout the attack with a Cosine Annealing learning rate scheduler \cite{loshchilov2022sgdr}. 

To measure the similarity between two sets of $N$-bit bitstreams we calculate the normalized Hamming distance as
\begin{equation}
    H(b_1, b_2) = \frac{1}{N}\sum_{n=1}^N \mathbb{I}(b_1(n) - b_2(n))
\end{equation}
where $\mathbb{I}(\cdot)$ is the indicator function  $\mathbb{I}(x)=1$ if $x\neq 0$ and 0 if $x= 0$. Then, given the Hamming distance, the attack success rate (ASR) is the percentage of adversarial images $x_{adv}$ whose compressed bitstream has a Hamming distance of 0 when compared with the compressed bitstream of the target image $x_{tgt}$. This can be written more formally as
\begin{equation}
    ASR = \frac{1}{M}\sum_{m=1}^M  \left(1-\mathbb{I} (H (b_{tgt}, b_{adv}^m)) \right)
\end{equation}
where $b_{tgt}$ is the compressed bitstream of the target image $x_{tgt}$ and $b_{adv}^m$ is the compressed bitstream of the $m$th adversarial image $x_{adv}$. 
We also tracked the quality of the adversarial images $x_{adv}$ using per-pixel $L_2$ norm
\begin{align}
    L_2(x, y) = \sqrt{ \frac{1}{M} \sum_{m=1}^M (x_m - y_m)^2}
\end{align}
where $x$ and $y$ are images consisting of $M$ pixels.

\begin{table*}
\centering
\begin{tabular}{|c|c|c|c|c||c|c|c||c|c|c|}\hline
  \multicolumn{2}{|c}{Attack Alg}& \multicolumn{3}{|c||}{MGD (ours)}& \multicolumn{3}{|c||}{PGD}& \multicolumn{3}{|c|}{CW}\\\hline \hline
     QF & NIC & CelebA & ImageNet & Kodak &  CelebA & ImageNet & Kodak & CelebA & ImageNet & Kodak\\ \hline
    & FP-GDN &   1.00 & 1.00 & 1.00 &     0 & 0 & 0 &    0 & 0 & 0 \\
   1 &  FP-ReLU &  1.00 & 1.00 & 1.00 &    0 & 0 & 0   & 0 & 0 & 0 \\
     & SH &  0.19 & 0.05 & 0.33   & 0 &  0 & 0  & 0 & 0 & 0 \\
     \hline
 & FP-GDN & 1.00& 1.00& 1.00& 0& 0& 0& 0 & 0   &0\\
 2& FP-ReLU & 1.00& 1.00& 1.00& 0& 0& 0& 0 & 0   &0\\
 & SH & 0.11 & 0.04 & 0.17 & 0 & 0 & 0   & 0 & 0 &0   \\
 \hline
 & FP-GDN & 1.00& 1.00& 1.00& 0& 0& 0& 0& 0&0\\
 3& FP-ReLU & 1.00& 1.00& 1.00& 0& 0& 0& 0 & 0   &0\\
 & SH & 0& 0& 0& 0 & 0 & 0   & 0 & 0 &0   \\
 \hline
 & FP-GDN & 0.32 & 0.94 & 0.92 & 0& 0& 0& 0& 0&0\\
 4& FP-ReLU & 0.32 & 0.94 & 0.92 & 0 & 0   & 0& 0 & 0   &0\\
 & SH & 0& 0& 0& 0 & 0 & 0   & 0 & 0 &0   \\
 \hline
  & FP-GDN & 0& 0& 0& 0& 0& 0& 0& 0&0\\
 $\geq$5& FP-ReLU & 0& 0& 0& 0 & 0   & 0& 0 & 0   &0\\
 & SH & 0& 0& 0& 0 & 0 & 0   & 0 & 0 &0   \\
 \hline 
 \hline

 any & JPEG & 0 & 0 & 0   & 0 & 0 & 0    & 0 & 0 & 0 \\ \hline

 \end{tabular}
        \caption{ASR of 3 attack algorithms (MGD, PGD, CW) applied to 3 NIC Models (FP-GDN, FP-ReLU, SH) plus a JPEG model over 3 datasets (CelebA, ImageNet, Kodak).}
    \label{tab:asr_model_attack1}
\end{table*}

\subsection{Performance of MGD Attack Algorithm}
\label{directattack}

In this set of experiments, we perturb source images $x_{src}$ and change them into adversarial images $x_{adv}$ that have the compressed bitstreams exactly the same as that of the target image $x_{tgt}$. The ASR values we obtained are shown in Table \ref{tab:asr_model_attack1}. We can easily see that bitstream collision is a real vulnerability to NICs. Our proposed MGD easily generated collision images with extremely high probability. In comparison, JPEG compression was extremely robust to bitstream collision attacks, as no valid collision image was obtained at all.

We also observe that NICs appeared robust against conventional attack algorithms PGD and CW. In other words, the proposed MGD attack was more powerful than PGD and CW.
SH was relatively more robust than FP-GDN and FP-ReLU. Some extra observations about the PGD performance are in Appendix \ref{pgdperf}.

\begin{figure}[t]
\centerline{\includegraphics[width=8cm]{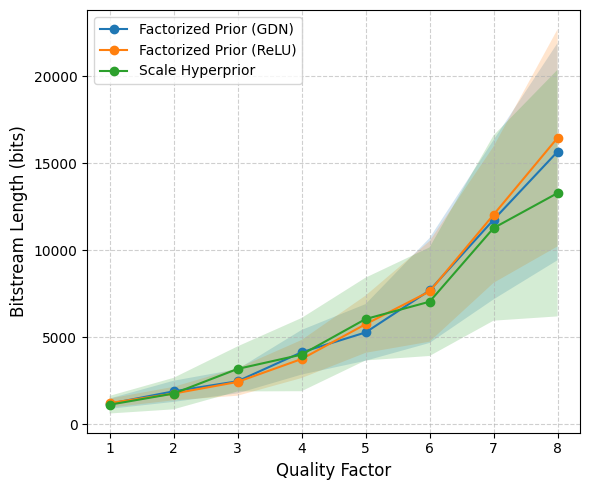}}
\caption{Compressed bitstream length vs Quality Factor (QF) over the three datasets. Note that bitstream length 10000 equals to compression ratio $R=150$, which is approximately the JPEG compression ratio of the datasets.}
\label{bitstream}
\end{figure}

There was a strong relationship between the quality factor (QF) of the compressor and the efficacy of the attack. As the QF increases,  the length of the compressed bitstream increases or the compression ratio decreases. This effect can be seen in Fig. \ref{bitstream}. For NICs to have higher compression ratio than JPEG, the QF should be less than $5$. But for small QF, bitstream collision becomes a severe vulnerability. 

Table \ref{tab:l2_model_attack1} shows the quality of the successful collision images $x_{adv}$ generated by our MGD algorithm, where $L_2(x_{adv}, x_{tgt})$ is the distance between $x_{adv}$ and the target image $x_{tgt}$, and $L_2(x_{adv}, x_{src})$ is the distance between $x_{adv}$ and $x_{src}$. We can see from the table that $L_2(x_{adv}, x_{tgt})$ was always very big, which means the adversarial images were perceptually different from the target image, even though their compressed bitstreams were the same. The $L_2(x_{adv}, x_{src})$ values were relatively smaller, indicating that the adversarial images were noise-perturbed versions from the source images. Extra perceptual data are shown in Appendix \ref{appendixssim} and image samples are shown in Appendix \ref{sampleimage}.

\begin{table*}
\centering
\begin{tabular}{|c|c|c|c|c||c|c|c|}\hline
   &\multicolumn{1}{|c|}{} & \multicolumn{3}{|c||}{$L_2(x_{adv}, x_{tgt})$} & \multicolumn{3}{|c|}{$L_2(x_{adv}, x_{src})$} \\ \hline
   QF&\multicolumn{1}{|c|}{NIC Model}& \multicolumn{1}{|c|}{CelebA}& \multicolumn{1}{|c|}{ImageNet}& \multicolumn{1}{|c||}{Kodak} & 
   \multicolumn{1}{|c|}{CelebA}& \multicolumn{1}{|c|}{ImageNet}& \multicolumn{1}{|c|}{Kodak}  
  \\\hline
  
   &FP-GDN &  0.88$\pm$0.12& 0.80$\pm$0.08& 0.71$\pm$0.08& 
0.62$\pm$0.04& 0.53$\pm$0.04& 0.57$\pm$0.08\\
  
   1&FP-ReLU &   0.90$\pm$0.14& 0.78$\pm$0.08,&   0.71$\pm$0.14&
0.68$\pm$0.08&  0.60$\pm$0.0.06&  0.61$\pm$0.10\\

   &SH &   1.89$\pm$0.32&  1.35$\pm$0.29&     0.98$\pm$0.16&
  
1.53$\pm$0.28&  1.01$\pm$0.19&    0.74$\pm$0.14\\
\hline

   &FP-GDN &  0.93$\pm$0.08& 0.89$\pm$0.12& 0.90$\pm$0.13& 
0.66$\pm$0.08& 0.62$\pm$0.04& 0.72$\pm$0.16\\
  
   2&FP-ReLU &   0.87$\pm$0.12& 0.82$\pm$0.14&   0.89$\pm$0.12&
0.68$\pm$0.09&  0.64$\pm$0.12&  0.71$\pm$0.12\\

   &SH &   1.80$\pm$0.34&  1.48$\pm$0.27&     1.36$\pm$0.24&
  
1.31$\pm$0.30&  1.02$\pm$0.18&    1.05$\pm$0.26\\
\hline

   &FP-GDN &  0.93$\pm$0.12& 0.93$\pm$0.08& 0.76$\pm$0.11& 
0.59$\pm$0.08& 0.61$\pm$0.05& 0.55$\pm$0.08\\
  
   3&FP-ReLU &   0.93$\pm$0.12& 0.86$\pm$0.16&   0.91$\pm$0.14&
0.62$\pm$0.08&  0.67$\pm$0.12&  0.70$\pm$0.18\\

   &SH &   -&  -&     -&
  
-&  -&    -\\
\hline

   &FP-GDN &  1.15$\pm$0.12, & 1.11$\pm$0.12, & 0.98$\pm$0.15, & 
0.87$\pm$0.08& 0.88$\pm$0.08& 0.80$\pm$0.13\\
  
   4&FP-ReLU &   1.06$\pm$0.12, & 0.99$\pm$0.14, &   1.08$\pm$0.11,&
0.89$\pm$0.11&  0.79$\pm$0.12&  0.80$\pm$0.10\\

   &SH &   -&  -&     -&
  
-&  -&    -\\
\hline
     \end{tabular}
        \caption{$L_2$ distance of {\bf successful} adversarial images $x_{adv}$ from $x_{tgt}$ or $x_{src}$. Each entry $a\pm b$ means average $L_2$ distance $a$ with standard deviation $b$. The dash '$-$' means no successful $x_{adv}$ found.}
    \label{tab:l2_model_attack1}

\end{table*}


    

\subsection{Transferability of MGD Attacks}
\label{trans}

We evaluated the transferability of adversarial images between NIC models. Specifically, we ran the attacks on a model to generate $x_{adv}$ that collided with $x_{tgt}$. Then, we checked if $x_{adv}$ still collided with $x_{tgt}$  on a different model. The results of these experiments are available in Tables \ref{tab:attack_transfer_celeba} and \ref{tab:attack_transfer_imgnet}. In the tables, the first column describes the models that the attack was conducted on to generate $x_{adv}$, and the first row are the models that $x_{adv}$ was evaluated on. Within the Factorized Prior family of models, the transfer attacks were essentially $100\%$ successful for each dataset that we evaluated on. While the attacks were transferrable within the Factorized Prior family of models, they were not transferrable between either of the Factorized Prior models and the Scale Hyperprior model. 

\begin{table}[t] 
    \centering
    \begin{tabular}{|c|c|c|c|} \hline
         NIC Model & {FP-GDN}& {FP-ReLU}& {SH}\\ \hline 
         {FP-GDN}& - & 96.7\%& 0\%\\ \hline 
         {FP-ReLU}& 95.6\%& - & 0\%\\ \hline 
         {SH}& 0\%& 0\%& - \\ \hline
    \end{tabular}
    \caption{Percentage of adversarial images that transfer between NIC models. CelebA dataset.}
    \label{tab:attack_transfer_celeba}
\end{table}

\begin{table}[t] 
    \centering
    \begin{tabular} {|c|c|c|c|} \hline
         NIC Model & {FP-GDN}& {FP-ReLU}& {SH}\\ \hline 
         {FP-GDN}& - & 90.3\%& 0\%\\ \hline 
         {FP-ReLU}& 89.9\%& - & 0\%\\ \hline 
         {SH}& 0\%& 0\%& - \\ \hline
    \end{tabular}
    \caption{Percentage of adversarial images that transfer between NIC models. ImageNet dataset.}
    \label{tab:attack_transfer_imgnet}
\end{table}

\subsection{Defense Performance of LPD}

We implemented the LPD defense to all the three NIC models (FP-GDN, FP-ReLU, SH) with quality factor $1$, and applied the proposed MGD attack. The ASR data are shown in Table \ref{tab:defense1}.
Prior to applying the defense, the attacks had a very high success rate, especially on the Factorized Prior models which had $100\%$ attack successful rate (Table \ref{tab:asr_model_attack1}). But our defense method reduced the ASR to 0 across all models and datasets that we tested. This demonstrates that the proposed defense method was extremely effective. 

\begin{table}[h] 
    \centering
    \begin{tabular}{|c|c|c|} \hline
    NIC Model &  CelebA   & ImageNet \\ \hline
    FP-GDN  & 0  & 0   \\ \hline
    FP-ReLU & 0  & 0    \\ \hline
    SH      & 0   & 0   \\ \hline
    \end{tabular}
    \caption{ASR after the LPD defense method is applied.}
    \label{tab:defense1}
\end{table}


\section{Conclusion}
\label{conclusion}

In this paper, we demonstrate that neural image compression (NIC) suffers from a unique vulnerability, bitstream collision, that conventional image compression algorithms such as JPEG do not. A new attack algorithm called Masked Gradient Descent (MGD) algorithm is developed to generate collision images that are perceptually different but have the identical compressed bitstreams. Surprisingly, this vulnerability may be mitigated by a simple Limited-Precision Defense (LPD) method, which implements some portions of the NIC models with limited resolution. Extensive experiments are conducted to verify the MGD algorithm and the LPD method. Considering the important role played by data compression in many security-critical applications, the collision vulnerability should be explored thoroughly before the real adoption of NIC technology.  




\bibliography{citations}
\bibliographystyle{icml2025}


\newpage
\appendix
\onecolumn
\section{MS-SSIM of Generated Images} \label{appendixssim}

Table \ref{tab:l2_model_attack1} shows the $L_2$ distance of the {\it successful} adversarial images. For a better representation of the perceptual similarity, the Multi-Scale Structural Similarity Index Measure (MS-SSIM) might be a more preferred performance measure. Table \ref{tab:ssim_model_attack2} shows the MS-SSIM of all the adversarial images, either successful or not, that we generated in our experiments. 
The data in the column of MS-SSIM$(x_{adv}, x_{tgt})$ are small in general, indicating that the adversarial images $x_{adv}$ were semantically different to $x_{tgt}$.  On the other hand, the data in the column of MS-SSIM$(x_{adv}, x_{src})$ are large in general, meaning the adversarial images $x_{adv}$ were semantically similar to $x_{src}$. In addition,
from the column of MS-SSIM$(x_{adv}, x_{src})$, the perturbation level of PGD and CW was similar as MGD but still failed (because the ASRs of PGD and CW were 0 in Table \ref{tab:asr_model_attack1}). Therefore, MGD is a more superior attack algorithm than the other two. 

\begin{table*}[h]
\centering
\begin{tabular}{|c|c|c|c||c|c|c|}\hline
  \multicolumn{1}{|c|}{} & \multicolumn{3}{|c||}{MS-SSIM$(x_{adv}, x_{tgt})$} & \multicolumn{3}{|c|}{MS-SSIM$(x_{adv}, x_{src})$} \\ \hline
  \multicolumn{1}{|c|}{Attacks}& \multicolumn{1}{|c|}{MGD (ours)}& \multicolumn{1}{|c|}{PGD}& \multicolumn{1}{|c||}{CW} & 
   \multicolumn{1}{|c|}{MGD (ours)}& \multicolumn{1}{|c|}{PGD}& \multicolumn{1}{|c|}{CW}  
  \\\hline
  
  FP-GDN &  0.14$\pm$0.04& 0.22$\pm$0.13& 0.41$\pm$0.08& 
0.75$\pm$0.11& 0.74$\pm$0.07& 0.69$\pm$0.06\\
  
  FP-ReLU &   0.11$\pm$0.03& 0.24$\pm$0.11&   0.43$\pm$0.08&
0.73$\pm$0.07&  0.75$\pm$0.08&  0.67$\pm$0.07\\

  SH &   0.18$\pm$0.03&  0.27$\pm$0.10&     0.40$\pm$0.06&
  
0.71$\pm$0.03&  0.75$\pm$0.05&    0.70$\pm$0.02\\

     \hline

\end{tabular}
        \caption{Average MS-SSIM of all the images $x_{adv}$, whether successful or unsuccessful collisions, for all the quality factors and all the three datasets.}
    \label{tab:ssim_model_attack2}
\end{table*}

\section{Further exploration of the effectiveness of PGD attack} \label{pgdperf}



\begin{figure}[htbp]
\centerline{\includegraphics[height=6cm]{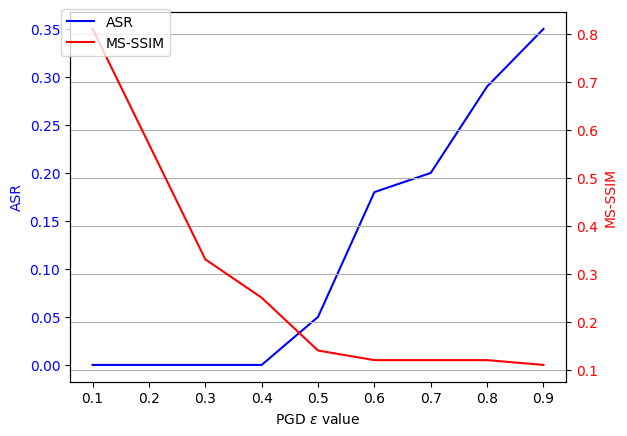}}
\caption{Evolution of ASR and MS-SSIM as $\epsilon$ changes in PGD}
\label{asr-mssim}
\end{figure}

When experimenting with the PGD attack algorithm to generate collision images in Section \ref{directattack}, we constrained the perturbations within a $\epsilon = 0.1$ ball to minimize distortion. However, this led to poor ASR for PGD across the board. We observed that increasing the magnitude of allowable perturbations caused the ASR to increase, but it also caused an increase of distortion. This effect is illustrated in Fig. \ref{asr-mssim}. When the ASR is increased from 0 to 0.05, the distortion of the adversarial image is already reduce to having MS-SSIM of 0.02 only. Note that the MS-SSIM is the Multi-Scale Structural Similarity Index Measure between $x_{adv}$ and $x_{src}$. As a result, the PGD attack algorithm is not competitive to the proposed MGD attack algorithm.

\section{Sample Adversarial Images}  \label{sampleimage}

\begin{figure}[htbp]
\centerline{\includegraphics[height=8cm]{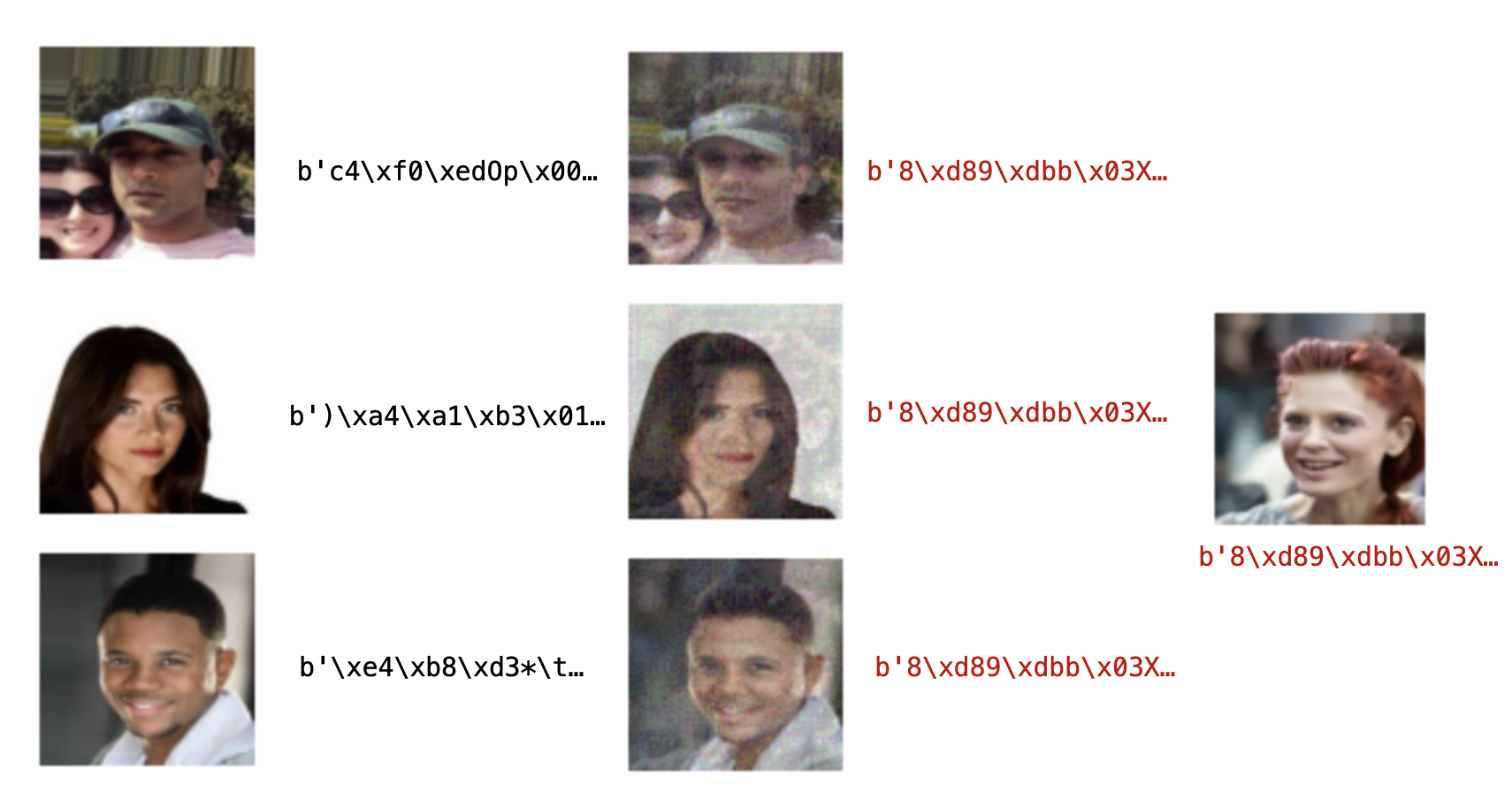}}
\caption{Images from the CelebA dataset with their corresponding compressed bitstreams generated with the Scale Hyperprior (SH) model.
Source Images and their compressed bitstreams (left). Adversarial Images and their compressed bitstreams (center). Target Image and its compressed bitstream (right).}
\label{sh_celeba}
\end{figure}

\begin{figure}[htbp]
\centerline{\includegraphics[height=8cm]{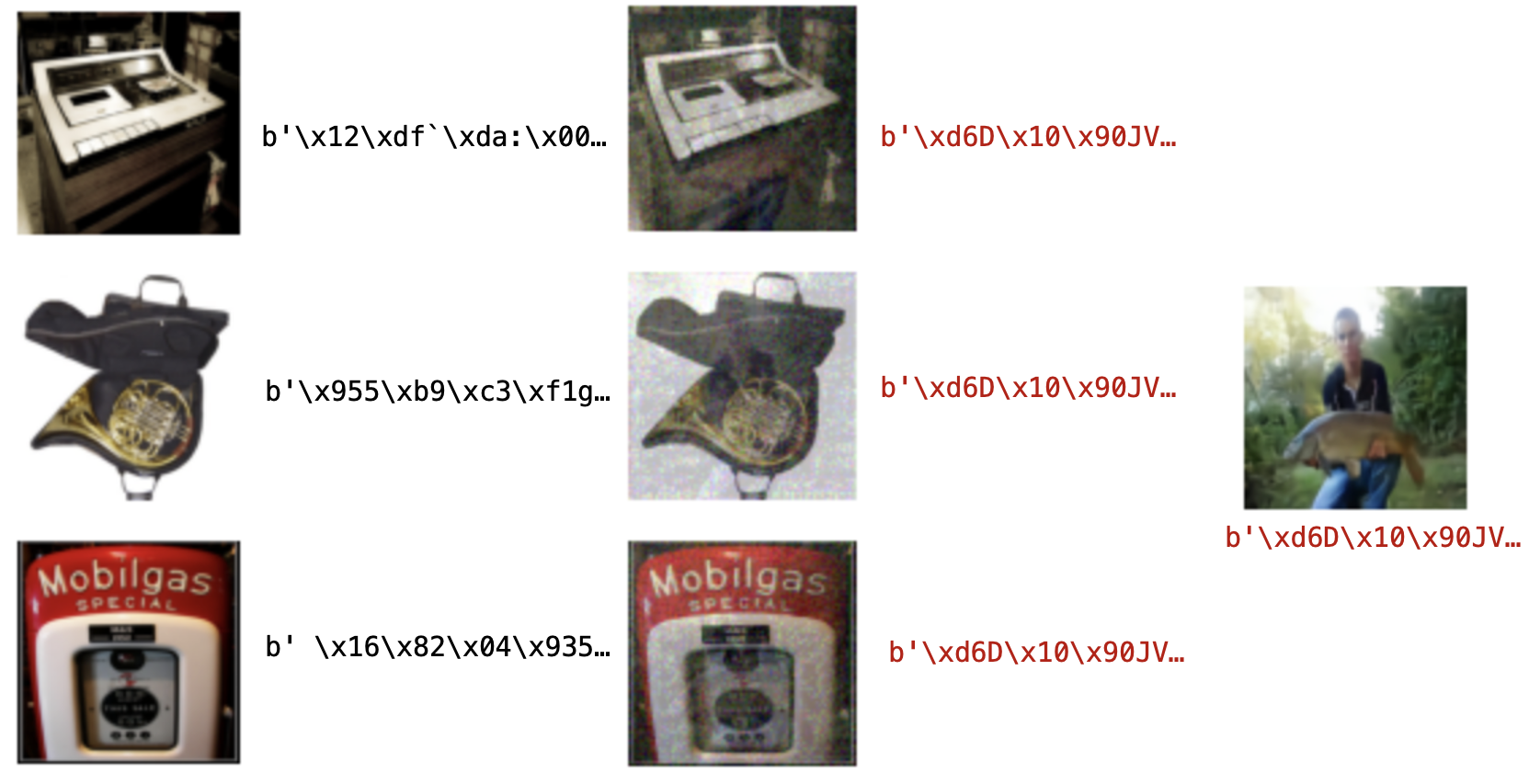}}
\caption{Images from the ImageNet dataset with their corresponding compressed bitstreams generated with the Scale Hyperprior (SH) model.
Source Images and their compressed bitstreams (left). Adversarial Images and their compressed bitstreams (center). Target Image and its compressed bitstream (right).}
\label{sh_imagenet}
\end{figure}

Figs. \ref{sh_celeba}-\ref{fp_gdn_kodak}
show samples of source images $x_{src}$, adversarial images $x_{adv}$, and target images $x_{tgt}$, as well as their compressed bitstreams. All the images $x_{adv}$ have bitstream collisions to the target images.

\begin{figure}[htbp]
\centerline{\includegraphics[height=8cm]{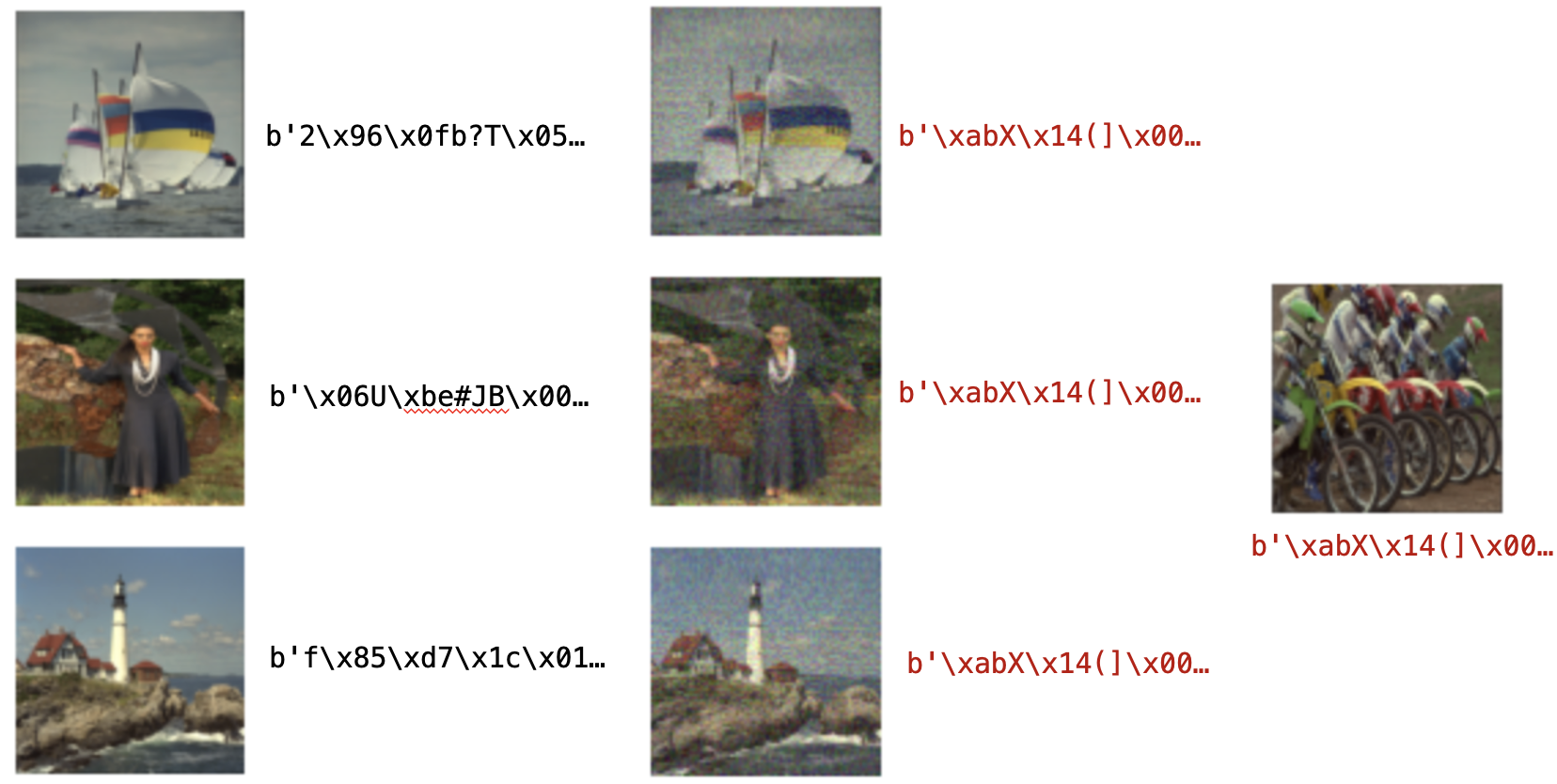}}
\caption{Images from the Kodak dataset with their corresponding compressed bitstreams generated with the Scale Hyperprior (SH) model.
Source Images and their compressed bitstreams (left). Adversarial Images and their compressed bitstreams (center). Target Image and its compressed bitstream (right).}
\label{sh_kodak
}
\end{figure}

\begin{figure}[htbp]
\centerline{\includegraphics[height=8cm]{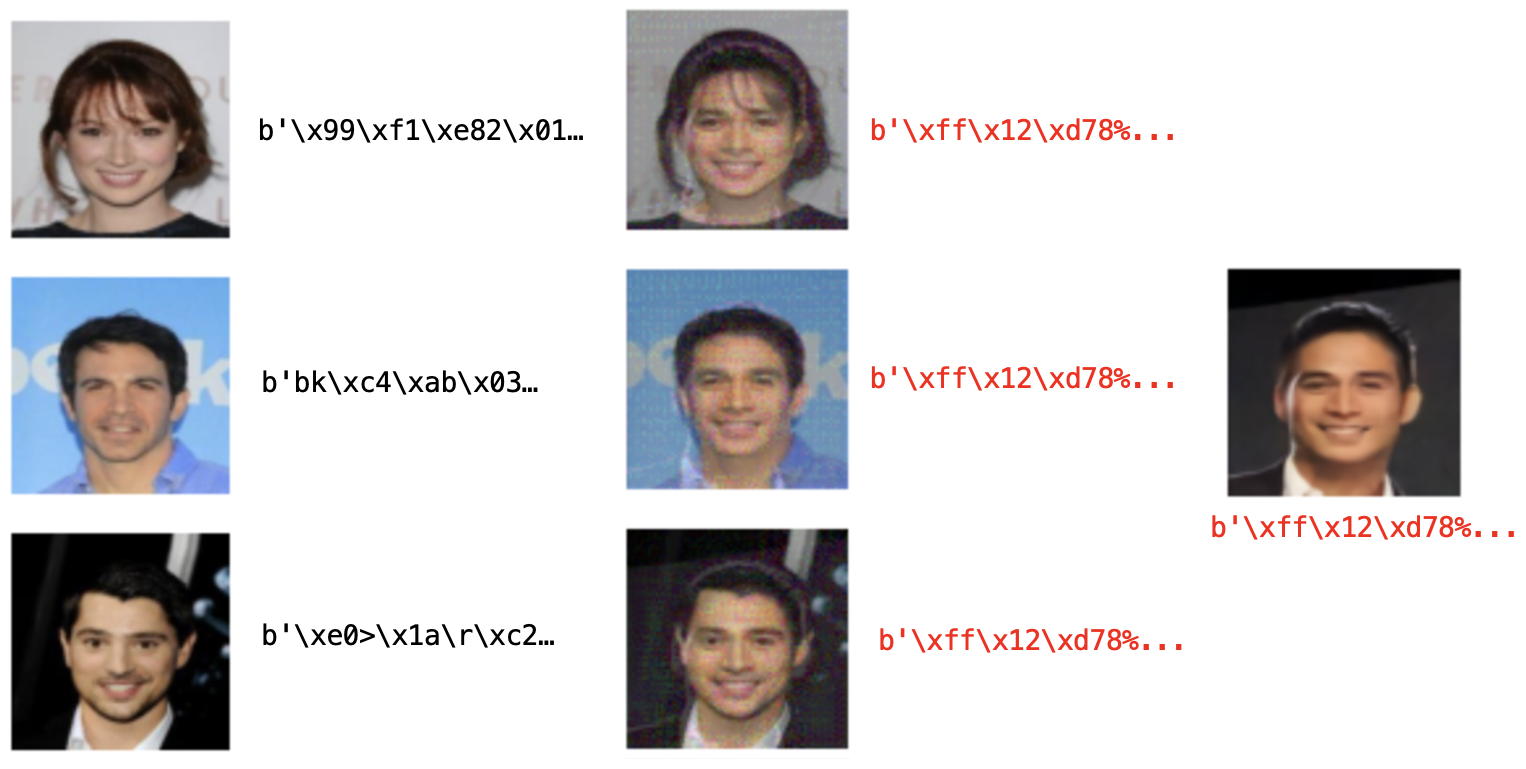}}
\caption{Images from the CelebA dataset with their corresponding compressed bitstreams generated with the Factorized Prior (ReLU) model.
Source Images and their compressed bitstreams (left). Adversarial Images and their compressed bitstreams (center). Target Image and its compressed bitstream (right).}
\label{fp_relu_celeba}
\end{figure}

\begin{figure}[htbp]
\centerline{\includegraphics[height=8cm]{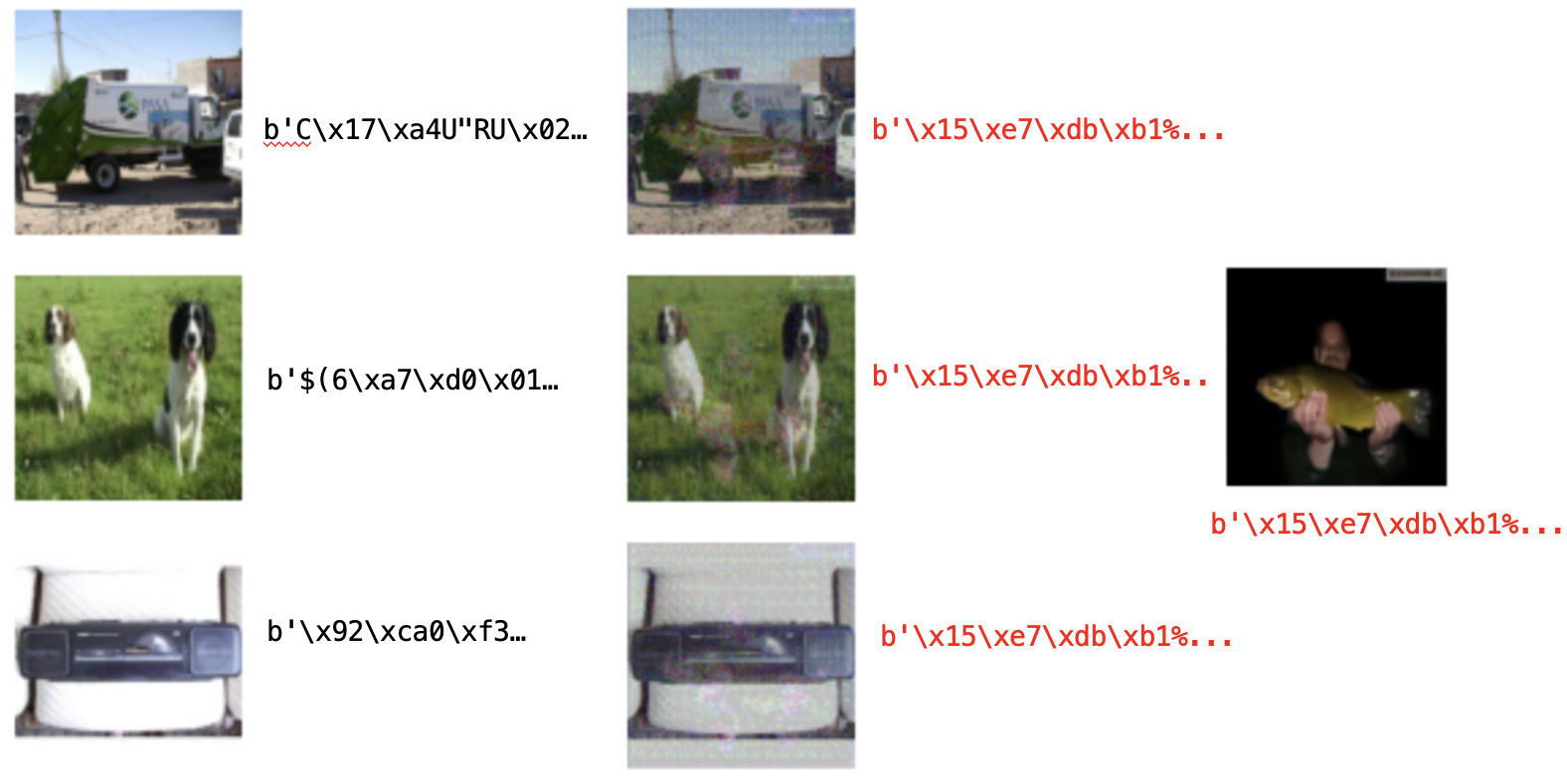}}
\caption{Images from the ImageNet dataset with their corresponding compressed bitstreams generated with the Factorized Prior (ReLU) model.
Source Images and their compressed bitstreams (left). Adversarial Images and their compressed bitstreams (center). Target Image and its compressed bitstream (right).}
\label{fp_relu_imagenet}
\end{figure}

\begin{figure}[htbp]
\centerline{\includegraphics[height=8cm]{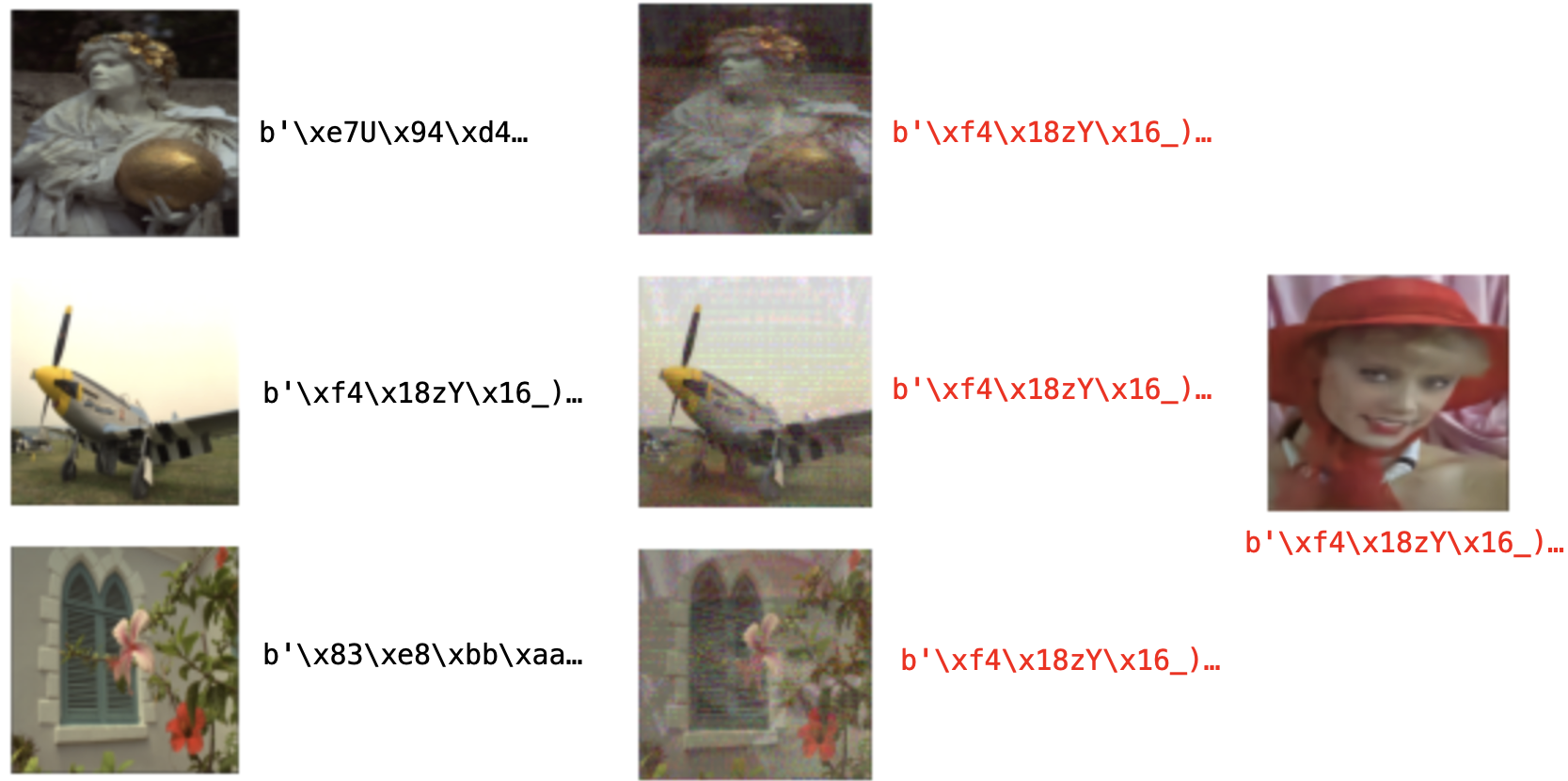}}
\caption{Images from the Kodak dataset with their corresponding compressed bitstreams generated with the Factorized Prior (ReLU) model.
Source Images and their compressed bitstreams (left). Adversarial Images and their compressed bitstreams (center). Target Image and its compressed bitstream (right).}
\label{fp_relu_kodak}
\end{figure}

\begin{figure}[htbp]
\centerline{\includegraphics[height=8cm]{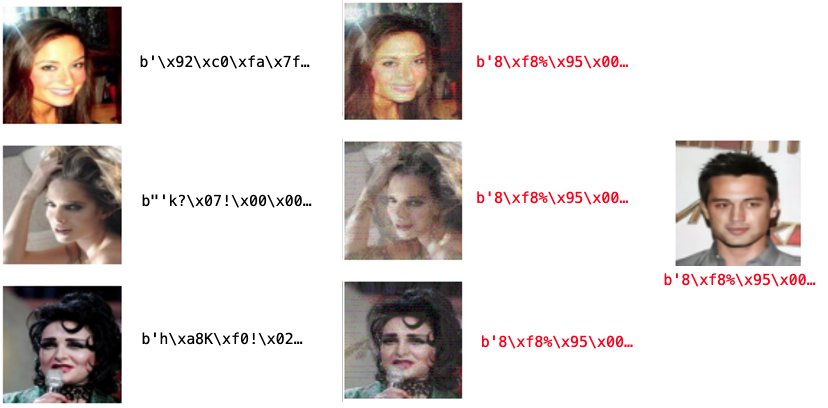}}
\caption{Images from the CelebA dataset with their corresponding compressed bitstreams generated with the Factorized Prior (GDN) model.
Source Images and their compressed bitstreams (left). Adversarial Images and their compressed bitstreams (center). Target Image and its compressed bitstream (right).}
\label{fp_gdn_celeba}
\end{figure}

\begin{figure}[htbp]
\centerline{\includegraphics[height=8cm]{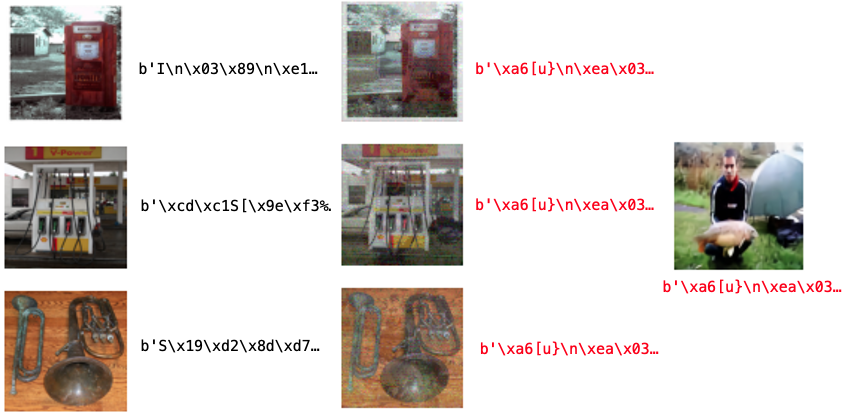}}
\caption{Images from the Imagenet dataset with their corresponding compressed bitstreams generated with the Factorized Prior (GDN) model.
Source Images and their compressed bitstreams (left). Adversarial Images and their compressed bitstreams (center). Target Image and its compressed bitstream (right).}
\label{fp_gdn_imagenet}
\end{figure}

\begin{figure}[htbp]
\centerline{\includegraphics[height=8cm]{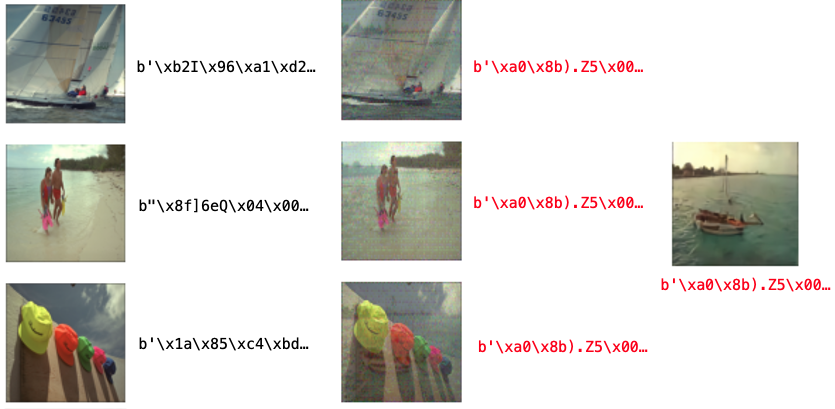}}
\caption{Images from the Kodak dataset with their corresponding compressed bitstreams generated with the Factorized Prior (GDN) model.
Source Images and their compressed bitstreams (left). Adversarial Images and their compressed bitstreams (center). Target Image and its compressed bitstream (right).}
\label{fp_gdn_kodak}
\end{figure}

\end{document}